\documentstyle[aasms4]{article} 

\newcommand{\kms}{{km s$^{-1}$}} 
\newcommand{\zabs}{{z$_{\rm abs}$}} 
\begin{document} 
 
\title{UVES observations of QSO 0000--2620:  oxygen and zinc abundances in the  
Damped Ly$\alpha$ galaxy 
at z$_{abs}=3.3901$
\footnote{Based on observations made with the ESO 8.2m KUEYEN telescope 
operated on Paranal Observatory, Chile.}  
}

\author{ Paolo Molaro$^1$,  Piercarlo Bonifacio$^1$, Miriam Centuri\'on$^1$,
Sandro D'Odorico$^2$, 
Giovanni Vladilo$^1$, Paolo Santin $^1$ and Paolo Di Marcantonio $^1$} 
\affil{$^1$ Osservatorio Astronomico di Trieste, Via G.B. Tiepolo 11, 34131, 
Trieste} 
\affil{$^2$European Southern Observatory, Karl
Schwarzschildstr. 2,D-85740 Garching, Germany}
\authoremail{molaro@oat.ts.astro.it}

	\begin{abstract} {Observations of  the QSO 0000--2620 
with UVES spectrograph at the 8.2m ESO KUEYEN telescope are used
for  abundance analysis of the damped Ly$\alpha$ system at
$z_{abs}$=3.3901. Several Oxygen lines are identified in the
Ly$\alpha$ forest and a  measure for the
oxygen abundance is obtained  at [O/H]=--1.85 $\pm$ 0.1 by means of 
the unsaturated OI $\lambda$ 925\AA\
and OI $\lambda$ 950\AA\ lines. This represents the most accurate O measurement 
 in a  damped Ly$\alpha$ galaxy so far.
We have also detected  ZnII $\lambda$ 2026 \AA\ and CrII $\lambda$2056, 
$\lambda$ 2062 \AA\
redshifted at $\approx$ 8900 \AA\ and found   abundances   [Zn/H] = --2.07$\pm$0.1 
and 
[Cr/H]=--1.99$\pm$0.1.
Furthermore, 
previous measurements of Fe, Si, Ni and N have been  refined yielding
 [Fe/H]=--2.04$\pm$0.1, [Si/H]=--1.90$\pm$0.1, [Ni/H]=--2.27$\pm$0.1,
and [N/H]=--2.68$\pm$ 0.1.
The abundance of the  non-refractory  element zinc is the lowest among the damped Ly$\alpha$
systems     showing
that  the associated intervening 
galaxy is indeed in the early stages of its chemical evolution. The fact 
that the Zn
 abundance is 
identical to that  of the refractory elements Fe and Cr 
suggests that dust grains have  not  formed yet.

In this Damped Ly$\alpha$ system  the observed [O,S,Si/Zn,Fe,Cr] ratios, 
in whatever combination are taken, are close to solar  
(i.e 0.1--0.2 dex) and
do not show the 
[$\alpha$-element/Fe] enhancement observed in Milky Way stars of
comparable  metallicity.   
The observed behavior 
supports   a   galaxy evolution model characterized by either episodic or low star
formation rate rather than  a Milky-Way-type evolutionary  model.}
 \end{abstract}

 
\section{Introduction}
 
Damped Ly$\alpha$ (DLA) systems ($\log$ N(HI) $\geq$ 20.3 cm$^{-2}$) are seen in absorption in 
the QSO spectra  
up to the highest redshift. They   likely trace 
the field population  of galaxies with no a priori selection  based on
morphology,
 since  biases apply   to the background sources at first approximation 
 and not 
to 
 the intervening systems. 
The identification of the galaxy population associated with the 
DLAs has important bearings on  cosmological elemental evolution and 
has been actively debated.

 Wolfe et al. (1995) and Prochaska \& Wolfe (1997,1999)  suggested
 a close connection between the
 DLA galaxies and the  progenitors of  
present-day spiral galaxies. The connection is based on   the fact that
the comoving gas density of the 
DLA galaxies at high redshift roughly equals the local density provided by
the luminous matter in nearby spiral galaxies. In addition,  the  kinematics
of the metal lines has  been  interpreted as  having formed in 
massive rotating disks. However, Rao \& Turnshek (2000) did not find a
decrease in the comoving gas density of DLA galaxies at low redshifts
as expected if the gas had been converted into stars,  and 
Haehnelt, Steinmetz, \& Rauch (1998) and 
Ledoux et al. (1998) did not find  the kinematical argument  very
compelling.

DLA systems  offer the  possibility to measure  abundances  accurately for a wide
range of elements  in a variety of normal, gas-rich galaxies spanning a 
considerable
look-back time.
The metallicity of  DLA galaxies 
is generally  inferred from the absolute abundance of the volatile 
element Zinc which  is   
found at $\sim$  10\% of solar, with 
 no clear evolution over the  
redshift interval from  $z_{abs}$ $\approx$ 0.5 up to 3 (Pettini et al. 1999).
The apparent lack  of evolution has been considered a problem for the 
 interpretation of DLAs as proto-spirals, 
 in particular at low redshift where abundances 
much closer to the
solar ones are expected. The models 
of Fritze et al. (1999) for Sa and Sd galaxies bracket the Zn 
data from the high
and low metallicity  sides  over the entire redshift range,
 but require that DLAs 
 are biased against early type spirals at low redshift.  A possible   bias 
 is represented by the  low gas and the high dust contents  
 of  early type galaxies at low
 redshift. 
  On the same line, Prantzos and Boisser (2000) found the 
observed Zn abundances consistent
 with a spiral origin once the empirical relation  18.8$\le$ [Zn/H]+$\log 
N(HI) \le$ 21
 is imposed to the models. Jimenez et al. (1999) found that
the population of Low Surface Brighness galaxies (LSB)
   better matches the observed zinc metallicities. 
  Imaging  of DLA galaxies at relatively low redshift revealed that
the population of DLA galaxies is not dominated by spirals but rather composed 
by a variety of  morphological types which include LSB, dwarf irregulars
and late type spirals.
 (Le Brun et al. 1997; Rao \& Turnshek 1998).
 
Relative elemental abundances are  an important complement to  the information
provided by absolute abundances such as zinc metallicity.
In particular,  ratios of elements  which are produced 
by Type I and Type II supernovae in different proportions provide 
 independent insights into defining
 the kind of chemical evolution.
Despite the efforts of different authors,   
it has been difficult to establish the [$\alpha$/Fe] abundance pattern
in DLA systems because of the different level
of depletion of  refractory elements onto dust grains.
In the compilation of Savaglio et al. (2000) the average of 37 
measures is $<$[Si/Fe]$>$
 +0.43 $\pm$ 0.18 \footnote{Using the customary  
definition [X/H]= log (X/HI) - log (X/H)\sun}.
At face value   this is consistent with what   is observed in  
 the Galactic halo stars, which  has been interpreted as evidence
for an origin of DLAs in proto-spirals (Lu et al. 1996, Prochaska \& Wolfe 
1999).
 
Evidence that  at least in some DLAs relative abundances  do not conform to 
those
of the Galactic halo stars was first provided by Molaro et al. (1996) in  
a first study  of the   abundances in the  DLA at $z_{abs}$=3.3901 towards 
QSO
0000--2620, which is considered again in this paper. 
Molaro, Centuri\'on, \& Vladilo (1998) and Centuri\'on et al. (2000)  
also found  the ratio   of the undepleted elements  Sulphur and Zinc    
 to be approximately solar   in a sample of half a dozen  DLAs. 
Vladilo (1998)  corrected  the abundances of Si and Fe
for the differential elemental depletion and obtained  
approximate  solar ratios ([Si/Fe]$\approx$0) in all cases
investigated, thus  suggesting  that solar ratios might be
rather common among DLAs. Pettini et al. (2000) considering  DLAs
with
low dust depletion  found
 [Si/Fe] ratios
 broadly in line with Galactic stellar values although there are also
examples of near-solar [Si/Fe] ratios at [Fe/H]$<$--1. Savaglio et al. (2000), 
after correcting for dust, found evidence for overabundance of a factor 2 
in Si  in about 1/4 of the DLA systems.

The [O/Zn]   ratio
is probably the best diagnostic tool for the
[$\alpha$/Fe]  ratio  we  have for DLAs since
Oxygen is produced by TypeII SNe  and   is  
essentially
undepleted in the interstellar medium.  Unfortunately, it is difficult 
to obtain  the oxygen abundance with  the required accuracy. In the present 
study of the z=3.3901 DLA system towards QSO 0000--2620
we 
present
an  accurate oxygen abundance determination  based on unsaturated 
lines,
 together with
a Zn detection  at the highest redshift ever obtained, and discuss
the [O/Zn] ratio in connection with
the chemical evolution  of the associated  galaxy.
 
\section{Observations and data reduction} 
 QSO 0000--2620 is a   bright  QSO (V=17.5) with
 z$_{em}$=4.108 discovered by C. Hazard, which shows a damped Ly$\alpha$ system
 at z$_{abs}$ = 3.3901. This  is one of the few damped  systems  
 at redshift
 greater than
 3 known so far and is  well suited for a detailed study of  chemical 
 abundances  in primordial galaxies. Measurements of abundances of
 this system have been obtained by Savaglio et al. (1994) and Molaro et
 al. (1996) from EMMI observations at the ESO NTT, and by Lu et
 al. (1996) and Prochaska $\&$ Wolfe (1999) from   HIRES observations at KeckI.
 
 A few spectra of QSO 0000--2620
were obtained as test observations during the first commissioning of  
the  Ultraviolet-Visual Echelle Spectrograph 
(UVES) on the 
Nasmyth focus 
of the ESO 8.2m KUEYEN  telescope at Paranal, Chile, in October 1999, which 
were released for public use.  In this paper we have used
the
data of higher quality: two exposures  of 4000 sec 
and one of 4500 sec secured on October 13, when the seeing, as given
by the telescope guide probe, was between 0.35 and 0.5 arcsec FWHM.
 The  observations were all obtained in  the standard dichroic $\#$2 mode 
 which includes
 the spectral region from 3700 to 5000 \AA\ in the blue arm 
and from 6700 \AA\ to 10500 \AA\ in the red arm.
 The CCD in the blue arm is a 2Kx4K, 15 
$\micron$ pixel size thinned, anti-reflection coated EEV CCD-44, while
in the red the detector is a mosaic of an EEV CCD-44
of the same type of the blue arm and a MIT/LL CCID-20.
 The average pixel scale in the direction of the dispersion is  0.22
and 0.16 arcsec/pixel for the blue and red arm, respectively. 
 The slit width was  
set at 0.9 arcsec. The CCDs were read-out in 2x2 pixel binned mode,
which gives a r.o.n. of 2.1 e$^{-}$ rms for the blue arm CCD,  and 2.1 
and 3.4 e$^{-}$ rms for 
the two red arm CCDs.  More details about  the instrument can be found
in Dekker, D'Odorico and Kaufer (2000).      
The full width at half maximum of the instrumental profile, $\Delta 
\lambda_{\rm instr}$, was measured from the emission lines of the 
Thorium-Argon lamp. The resulting resolving power is R= $\lambda$/$\Delta 
\lambda_{instr}$  $\simeq$ 48000, which
 corresponds to  
a velocity  resolution of $\simeq$ 6 \kms.

The data reduction was performed using the ECHELLE context routines 
implemented  in the ESO MIDAS package.     
Flat-fielding, cosmic ray removal, sky subtraction,  and
wavelength calibration 
were performed  on each spectrum  separately.
Typical r.m.s  of the wavelength calibrations are 
$\leq$0.6 \kms.  
  
The observed wavelength scale of each spectrum was then
transformed into vacuum, 
heliocentric wavelength scale. The  spectra were then 
added together by  using  their S/N as  weights.   
Finally,  the local continuum was determined
in the average spectrum by using a spline to smoothly connect the  
regions free from absorption features.  The continuum for the Ly$\alpha$ 
forest region
was fitted by using the small regions deemed to
be free of absorptions and by interpolating between these regions with
a spline.
 
With respect to the previous ESO NTT observations, UVES spectra
have a higher extension into the blue  as well as a  more than double 
  resolution and 
higher S/N. With respect to the Keck data, which
cover 
the range 
5100--7660 \AA, UVES spectra extend further towards both the blue and the red 
sides. In the region
of overlap, which extends from $\sim$ 6700 to 7660 \AA\, 
the data have   similar resolution
and comparable S/N as can be seen from a comparison of our figures with
Fig.1 of Prochaska $\&$ Wolfe (1999).\\

\section{Measurements of the Column Densities}
 
In this work we  restricted our  analysis to the  metal lines
falling in the Ly $\alpha$ forest which have  never been observed at both comparable resolution
and S/N, and to lines in the reddest part of the spectrum. 
Measurements are presented in Table 1.
Column densities have been obtained by
fitting theoretical Voigt profiles 
to the observed absorption lines   
via $\chi^2$ minimization. 
The fit  was performed   using the  
  FITLYMAN package  within MIDAS (Fontana \& Ballester 1995). During 
  the fitting procedure the theoretical profiles were convolved    
with the instrumental point spread function 
modeled from the analysis of the emission lines of the arcs.  
Portions of the profiles recognized as contaminated 
by intervening Ly $\alpha$ clouds
were excluded from the analysis.  

The FITLYMAN routines determine
the  redshift, the column density, and the
broadening parameter ($b$-value)\footnote
{The broadening parameter is defined as $b = 2^{1/2} \sigma_v$,
where $\sigma_v$ is the one-dimensional gaussian velocity dispersion of ions 
along the line of sight. 
} of the 
absorption components, as well as the fit errors for 
each  quantity. 
The atomic data  were taken  from the compilation of Morton (1991) with the revisions 
by 
Cardelli \& Savage (1995), Bergeson \& Lawler (1993a,b) and Fedchak and Lawler (1999),
as specified in Table 1.
 
 Since Ly$\alpha$ is not
 included in our range, when calculating the abundances 
  the hydrogen column density is taken 
 as $\log N(HI)$=21.41$\pm$0.08 from Lu et al. (1996), 
  after having  verified  the value  consistent with
 the Ly$\beta$ absorption.
 With such a large column density, 
  the ion  species   are from
 the dominant ionization stages in HI gas and  do  not require  ionization 
 corrections 
 (Viegas 1995). The damped structure is known to include   minor
 additional components on both sides of the main absorption (Molaro et al. 
1996, Lu et al. 1996),
 but  our analysis is restricted to the unsaturated lines were these weak  
components 
do not appear  and do not affect the abundances. 
  
The derived  
column densities and $b$-values, together with the redshifts,
 are reported in Table 1.  
The difficulty in  determining accurate OI column densities stems from the fact 
that 
the only transition available redwards the Ly$\alpha$ emission is  OI
$\lambda$ 1302\AA\ which  has always been found heavily saturated in DLAs.
On the other hand  OI $\lambda$ 1356 \AA\
 has never been   detected and has provided  only 
loose  upper limits.  
Upon searching the Ly$\alpha$ forest we detected
 several oxygen lines, namely 
OI   $\lambda$925.446, $\lambda$929.517,  $\lambda$ 936.629,  $\lambda$ 
948.685,  
 $\lambda$950.884,  $\lambda$988.773 , $\lambda$ 1039.230 \AA\, 
 which are relatively free 
from 
contamination caused by   hydrogen clouds.  
The OI $\lambda$ 925 \AA\ and  OI $\lambda$950 \AA\ lines
 have a very low  {\it f} value 
 (3.5$\times^{-4}$
 and 15.7$ \times ^{-4}$), which is  140 and 25 times lower than the OI
$\lambda$ 1302 \AA\~, respectively. The lines   are not saturated  
thus  allowing accurate 
 determination of the oxygen abundance. The two oxygen 
 lines are shown in Fig. 1. 
Owing to the  presence of minor contaminants 
we 
analyzed the  two lines independently.  The two lines yield similar column
 densities, where that
obtained from  OI $\lambda$ 925 \AA\ is slightly higher. We notice that the 
  OI $\lambda$ 925 \AA\  is somewhat broader ($b$=14 km s$^{-1}$) than  
   the other
metal  lines, thus  suggesting the presence of
 some blend which may explain the small difference in the column densities. 
In fitting   OI $\lambda$ 950 the presence of the wing of the Ly$\delta$  
line on 
the blue side is accounted for. The adopted  abundance for OI is taken as  the 
straight average of the 
two measurements, i.e. $\log$N(OI)=16.42 (or [O/H]=--1.85$\pm$0.1) 
with an error of $\pm$0.1
which accounts for the uncertainty  in the continuum placement.
The other $\alpha$-element, Silicon, is obtained from the 
unsaturated SiII $\lambda$ 1808 \AA\  line also shown in Fig. 1.
We obtained a  column density of $\log$ N(SiII)=15.06, which is  in 
 good agreement with the value of 15.086 
 derived by Prochaska \& Wolfe (1999).

Several N transitions   belonging to  the NI $\lambda$952, $\lambda$953, 
$\lambda$956, $\lambda$1134 \AA\ multiplets have been 
 detected for the first time 
in a DLA. In each multiplet, however, only few lines  resulted 
relatively free 
from contaminations, namely NI $\lambda$ 952.415, 
$\lambda$ 953.415, $\lambda$ 953.655, 
NI $\lambda$ 963.990,  and $\lambda$ 965.041 \AA\ .
In deriving the nitrogen abundance we further restricted the analysis to 
those lines which appear 
totally free from  contaminations, which comprise the  three  lines of the
NI $\lambda$ 1134 \AA\ multiplet and the   
NI $\lambda$ 953.6 \AA\ transition and are   shown in Fig 2. 
The N column density is found
$\log$ (NI) =14.70, which is 
 in excellent agreement with the $\log$ N(NI)= 14.68 $\pm$ 0.14 
derived by Molaro
et al. (1996) from the analysis of the NI $\lambda$ 1134 \AA\ multiplet. 
   
The iron abundance is obtained from the unsaturated FeI $\lambda$ 1611 \AA\
line. The
abundance is determined at  [Fe/H]=--2.04, which is 0.3 dex  higher than the 
value obtained 
by 
Molaro et al. (1996), and 
0.2 dex lower than that obtained by
Prochaska and Wolfe (1999) by means of  the same transition.
The FeII line together with the other iron-peak elements are shown
in Fig. 3. 
 
Chromium is clearly detected through the CrII $\lambda$ 2056 \AA\
and $\lambda$ 2062 \AA\ transitions,
 while the CrII $\lambda$ 2066 \AA\ line is obscured by a 
strong sky emission line. The two lines yield almost identical Cr column 
densities with a  mean value of  $\log$N(CrII)=13.10. 

Zinc abundance is obtained from the stronger component of the 
doublet Zn II $\lambda$ 2026.1360 \AA\, while the weaker 
ZnII $\lambda$ 2062.664 \AA\ remains below the detection threshold.
The value found here is [Zn/H]=--2.07 which 
is  consistent with the  upper limits 
set at [Zn/H]$\le$--1.76 (3$\sigma$) by Pettini et al. (1995).

We have clearly detected the NiII $\lambda$ 1709 \AA\, while  
the NiII $\lambda$ 1751 \AA\ showed to be partially blended
with an O$_{2}$ atmospheric line. The abundances of  NiII 
$\lambda$ 1709 \AA\ yields [Ni/H]=--2.27,
which is also consistent with the value of  NiII $\lambda$ 1751 \AA\ when the 
telluric contamination
is accounted for. We note that the abundance of Ni, despite the recent upwards 
revision in the $f$ values by Fedchak and Lawler (1999), remains slightly 
lower than the other iron-peak elements.

Other transitions present in our spectrum  such as  AlII and CIV  are 
either saturated  or are 
 not significantly different from the values published in the quoted
references. In this paper they are not examined again.

\section{Discussion}

\subsection{ Zinc Abundance and Dust Content}

Zinc abundances  have been used to trace back the cosmological  chemical 
evolution
of Damped Ly$\alpha$ galaxies. Instead of the expected growth of metallicity
 with cosmic time, which is a  natural
outcome  of stellar evolution,
Pettini et al. (1999) could not find  any trend of the 
[Zn/H] ratio with z$_{\rm abs}$.
 The DLA towards QSO 0000--2620  is the only one at
 z$\ge$ 3  for which  a Zn abundance measurement is 
available at present.
The  zinc abundance  is found at 
[Zn/H]=--2.07, which  is the lowest level found  so far in any DLA at 
any redshift.
The low  metallicity level   shows 
that  this 
Ly$\alpha$ galaxy is indeed in the early stages of its chemical evolution. 
When  compared with   extant data, this  Zn abundance 
suggests that a mild chemical evolution is
present  in the DLA population.  Observational biases may affect 
this conclusion, as discussed
 in a separate paper (Vladilo et al. 2000).

A second important point  is that  zinc abundance at [Zn/H]=--2.07 
is fully  consistent with both iron and chromium abundances, 
which are [Fe/H]=--2.04 and 
[Cr/H]=--2.0
respectively. It is well established that  
 in the  DLAs Zn is consistently  more abundant than Cr and Fe  
 (Pettini et al. 1994, 1997), while in Galactic halo stars
 both Cr and Zn track the Fe content down to very low metallicities.
The different behavior of Fe  and Zn  in the DLAs is  currently
interpreted as the result of  differential 
depletion of these  elements from the gas phase onto dust grains 
 in analogy with the nearby interstellar medium.
The fact that in this DLA the abundances of the refractory elements
Fe and Cr are found close
to the non-refractory element Zn suggests 
that dust grains have not yet 
formed in this
protogalaxy. Such a circumstance is extremely  rare among  DLAs, and even
in 
the  cases 
with low depletion considered by Pettini et al. (2000),
 a difference between Zn 
and the other refractory
elements Fe, Cr and Ni has been always  observed. The absence of dust 
in this damped system
frees the measurements
of relative abundances from  the uncertainties deriving from the unknown
 fraction of elements which condensate onto grains, thus making this DLA of
particular value in the study of chemical composition patterns.
 
\subsection{Abundances of $\alpha$-capture versus iron-peak elements}

Information on $\alpha$-capture elements is important because 
in the early stages of the chemical evolution of galaxies the abundances 
are likely dominated by  
Type II SNe  products which are richer in $\alpha$-elements 
than Type I SNe  which entered into the game
 only later on. Thus in the early stages  the $\alpha$-capture elements  
are expected to  be relatively more abundant than the  iron-peak elements, 
and the
ratio should reverse later on during evolution.
 The exact  timing of the turn-over depends on both the SFR and IMF,  which 
differ for the different types 
of galaxies. The [$\alpha$/Fe] ratio is therefore an  indicator
 of the chemical evolution history which can be used to understand the
nature of galaxies when  data on their morphology and colors are lacking. 
Silicon is the most  accessible $\alpha$ element  in DLAs and  
 the compilation of Savaglio et al. (2000), who consider
 37 measures, obtain a mean value of
$<$[Si/Fe]$>$=0.43$\pm$0.18.
The problem here is that   we 
observe 
a similar ratio  ([Si/Fe]=+0.36) in the Warm Phase of 
 the   Galactic interstellar medium, due to the fact that the gaseous  silicon 
and iron abundances
are  lowered  by different  amounts
by dust depletion (Savage \& Sembach 1996). 
After accounting for  dust, Savaglio et al. (2000) conclude that Si is 
intrinsically
enhanced in comparison  to iron
only in about 25\% of the DLAs with SiII determined.
 Outram et al. (1999) in two DLA systems inferred 
  the absence of dust from the similarity of Si 
and S abundances and argued for 
an intrinsic overabundance of 
[$\alpha$/Fe]
of about 0.5 dex. 
 However, as noted in Matteucci et al. (1997),  Si might  be produced in Type 
I SNe, which could  explain the complicate observational pattern. Therefore, 
Si should be considered an   indicator of $\alpha$ element abundances
 less reliable than  O or S.

Evidence  that relative abundances in some DLAs do not conform to those
of the galactic halo stars has been provided by the [S/Zn] ratio
(Molaro et al. 1996, 1998,  Centuri\'on et al. 2000)  in a  sample of 6  DLAs
 but with one  case, the DLA at \zabs=2.476 towards Q0841+129,
  consistent with
[S/Zn]$>$0.2.

The oxygen abundance we derive here is the first one obtained by means of
unsaturated lines. Since the transitions lie in the forest,
the uncertainty is dominated by possible  Ly$\alpha$ interlopers.   
However, if these hydrogen lines do actually 
contaminate the spectrum at the wavelength of the oxygen
lines, 
the real  oxygen abundance should be even lower  than the  value 
found here.

Oxygen is a typical  product of type II SNe  and in the
atmospheres of Galactic metal-poor stars it shows the notorious
 enhancement  of   
[O/Fe] $\simeq$ +0.5. 
Claims for  an even more extreme
oxygen overabundance    with  [O/Fe] reaching 
+1.0 dex at [Fe/H]=--3.0 have been made recently by 
Israelian et al. (1998)  and Boesgaard et al. (1999).

Oxygen has a rather small condensation temperature (T$_{c}$ $\approx$ 180 K) 
and its depletion factor is of  
0.00$^{+0.18}_{--0.31}$ dex in the warm gas phase towards $\zeta$ Ophiuchi 
(Savage \& Sembach 1996). Therefore, 
the [O/Zn] ratio  is an excellent  dust-free diagnostic tool for   
the [$\alpha$/iron-peak] abundance ratio.
  
 From the average of the two transitions we derive [O/H]=--1.86,  which
 implies [O/Zn]=0.21,
 a relatively modest, if any,
 $\alpha$ enhancement. 
  Si abundance  is almost  the same as  oxygen and  we obtain [Si/H]=--1.91,
which   corresponds  to [Si/Zn]=0.16.   We note also here that
 Lu et al. (1996) tentatively identified 
the SII 1253.811 \AA\ line associated with the damped, and derived
a column density $\log$ N(SII) = 14.70 $\pm$ 0.03, i.e.  an abundance
 of [S/H]=--1.91, which implies [S/Zn]=0.16. Thus,
 the S abundance of Lu et al. (1996) is also consistent with  our O and Si 
 abundances in the system.

 For the first time in the DLA studied here we have at our disposal 
 a set of 3 indicators for the $\alpha$
 element abundances and
 three  indicators for the iron peak elemental abundances, which do
 not
 need any dust
 correction.  
 The [O,Si,S/Zn,Fe,Cr] ratios, in whatever combination the elements are 
chosen, 
are   in the range 0.09--0.22 dex. This range of values is  significantly 
lower than analogous ratios in Galactic stars with  comparable metallicities, 
and the difference is more stringent in this case for the
low value of the metallicity, which is one order of magnitude lower than 
the average  value in DLAs.
The lack of significant [$\alpha$/Fe] enhancement  corroborates previous 
results on this system 
as well as  the suggestion that,  on the basis of their relative elements,
at least some  DLA galaxies seem to undergo a chemical evolution which
differs from that of  the Milky--Way. 
Centuri\'on et al. (2000)  showed that the [S/Zn] ratios in DLAs 
are consistent with 
a decreasing trend  with increasing 
metallicity. These ratios show also  a correlation, 
although less clear-cut, 
 with cosmic time, where   the lower [S/Zn] values  
occur  mostly at low redshifts.
 The [S/Zn]=0.16 and [O/Zn]=0.19 for our DLA 
add a new point which supports the presence of such  trends. 
Considering the extant data of  [S/Zn] and the value [O/Zn]=0.20 we 
derived here, 
the regression analysis yields the correlation [S,O/Zn] =
 --0.36($\pm$0.11)[Zn/H] --
0.52. 
The $\alpha$/iron-peak ratio becomes  solar  
at [Fe/H] $\approx$-1.4 and further decreases  at higher 
metallicities.

Chemical evolutionary models predict   solar  [$\alpha$/Fe] at 
low  
metallicities,  when star 
formation proceeds in bursts separated by relatively long quiescent periods
or whenever star formation  
is slower than that  of our Galaxy. These conditions are met in 
 dwarf galaxies,  LSB galaxies and  in the 
outer regions of disks (Matteucci et al. 1997, Jim\'enez et al. 1999). 
In these galaxies
the metal enrichment is so slow that 
Type Ia supernovae have enough time to evolve and enrich the gas 
with iron-peak elements,    
in such a way that  reduced $\alpha$ over iron ratios are 
attained at low metallicity.

In the Milky Way  few metal poor stars with
solar [$\alpha$/Fe-peak] ratios have been found
(Carney et al. 1997, Nissen \& Schuster 1997). These  exceptional  stars
show large apogalactic distances  and  it was  suggested that 
they  belonged
to a satellite galaxy which  experienced a different chemical evolution
history than the Milky Way, and were subsequently accreted by the latter.

Incidentally, we note that the  interpretation of   the  low [O/Fe] ratio
 at low metallicities as an intrinsic  effect of TypeI SNe  
 does not support  the suggestion  by
 Umeda et al.  (1999) that  Type I SNe may be suppressed
 at metallicities lower that [Fe/H] $<$--1.0. 
 
\subsection
{The Nitrogen abundance}

In our system the  nitrogen abundances  relative to Zn and O are [N/Zn]=--0.54 
and
  [N/O]=--0.74. Both values are somewhat lower than previously 
 reported by Molaro et al. 
(1996)  
because of the upwards revision in  oxygen and iron abundances.
 The [N/O] is not 
so extreme as previously thought, but the N abundance still requires a primary production
of the element according to the  evolutionary models discussed
in Matteucci et al (1997).  N abundance  in this DLA, but  
also in other DLAs,  remain lower than the 
corresponding 
values in the 
Milky Way, where N is found to trace Fe in lockstep  (Israelian et al. 2000),
  which again suggests a different pattern than the Milky Way. 
 Nitrogen abundances have  been  discussed by
  Lu, Sargent, \& Barlow (1998) and by  
Centuri\'on et al. (1998),  who pointed out   the presence of   a  real scatter  
among  DLA systems
with some values very close to a pure secondary behavior  and 
other values
which require  a primary nucleosynthesis. The different values of 
N/O observed at a given
O/H may be understood in terms of the delayed delivery of primary N with respect to O 
when star formation proceeds in bursts.
During the quiescence period N is deposited in the ISM and the N/O ratio
increases while O/H remains constant. Therefore a relatively high [N/O], such 
as that 
observed here,   nicely fits with the picture of a delayed N release after 
a quiescence
period, which is also required by the low $\alpha$ over iron ratios.

\subsection{Conclusions}
 
UVES observations of the QSO 0000-2620 resulted in the detection 
of the ZnII 2026.136 \AA\ transition originated in the damped system at
z$_{abs}$=3.3901. The abundance derived is  [Zn/H]=-2.06 which is presently the
lowest  among  DLAs. This low  metallicity level shows  that the galaxy 
is in the early
stages of its chemical
evolution. When compared with the larger abundances observed in DLA galaxies at lower
redshift, this measurement
provides a first hint of a cosmological chemical evolution in which
abundances increase with decreasing redshift (see Vladilo et al 2000)

The abundances of Cr and Fe were also obtained and  found  similar
to the abundance  of Zn. This coincidence in the abundances between 
refractory and non refractory elements
is interpreted as   dust has  not  formed yet. 

The OI 925 \AA\ and 950 \AA\ lines were detected 
although they fall within the Lyman$\alpha$ forest. These lines have oscillator
strengths  much smaller than that of the OI 1302 \AA\ line 
which is strongly 
saturated. These detections allow, for the first time, a rather 
accurate measurement of the 
oxygen abundance in a DLA system.   The oxygen abundance, [O/H]=-1.85, is 
remarkably similar to  the sulphur 
abundance derived by Lu et al (1996), [S/H]=-1.91, and both 
abundances are similar
to that of silicon, [Si/H]=-1.91, which is consistent with the absence of
dust in the system as implied by  Zn and Fe abundances.

In this DLA  the relative abundances of $\alpha$ and iron-peak 
elements are  mildly overabundant compared to solar 
and  do not show the 
 enhancement  expected for  a  progenitor of  an early type spiral such as our own Galaxy. 
The paradigm of DLAs as  progenitors of the Milky-Way type spirals
 is not supported by  
the  study of the dust-independent chemical evolution indicators 
  such as  [O/Zn] and [S/Zn]. 
  The almost solar  abundance ratios  measured at the low   metallicity of
 [Fe/H]$\approx$=--2
suggest that 
this DLA should be associated with objects with low, or episodic, star 
formation rates
such as LSB or dwarf galaxies.  However, these results are not 
characteristic of this DLA alone and  similar low [S/Zn] ratios are also found 
  in other DLAs (Centuri\'on et al 2000).

\section{Acknowledgements}
The new high resolution spectra analyzed in this paper are of unique
quality
and were obtained during the first  nights  of 
commissioning of a new instrument at a new telescope. For these results
we are indebted
to all  ESO staff involved in the VLT construction and in 
UVES commissioning.

\clearpage

Figure caption:
\\
\\

\figcaption[fig1.epsi]{Observed lines for the $\alpha$-elements O and Si 
associated with  the 
z$_{abs}$=3.3901 DLA system towards  QSO 0000--2620.  All the lines
 are aligned taking zero velocity
in correspondence of z$_{abs}$=3.39010.
Smooth lines are the  synthetic spectra obtained from the fit  
as described in the text. The dotted line in the bottom panel bluewards the
SiII feature is the  CIV $\lambda \lambda$ 1548 \AA\ 
of the z$_{abs}$$\approx$ 4.126 metallic
system. The absorption features in the red side are the 
CIV $\lambda \lambda$ 1548 \AA\ 
of the z$_{abs}$ $\approx$  4.130 and 4.131. 
Cosmic removals are marked  by crosses.
\label{fig1}}

\figcaption[fig2.epsi]{As Fig 1 for the nitrogen lines. The top panel
is centered on the NI  $\lambda \lambda$ 1134.4149 \AA\ transition, with the
other  two lines NI $\lambda \lambda$ 1134.1653 \AA\ and NI $\lambda \lambda$ 
1134.9803 \AA\ on the bue and red sides, 
respectively.
\label{fig2}}

\figcaption[fig3.epsi]{As Fig 1 for the iron-peak elements. The two dotted features
on both sides of CrII $\lambda \lambda$ 2056.2539 \AA\ are telluric lines.
\label{fig3}} 
\clearpage


 

\pagestyle{empty}
\begin{deluxetable}{lcccccccc}
\footnotesize
\tablewidth{-5cm} 
\tablecaption{Column densities.} 
\tablehead{
\colhead{ Ion }   & \colhead{$\lambda$}&
\colhead{f $\tablenotemark{a}$}  &\colhead{$\log$ (N)} & \colhead{b} & 
\colhead{$z_{abs}$ }  & \colhead{(Z/H)$_{\sun}$} &
\colhead{[Z/H]} & \colhead{[Z/Zn]}
} 
\startdata
HI    & 1215.670   &   0.4164    &  21.41 $\pm$0.08 $\tablenotemark{b}$ &     
&   &  &  & \nl
\nl
NI    & 953.6549      & 0.02032   & 14.77$\pm$0.03  & 9.4$\pm$0.7  & 3.390163
        &  --4.03$\pm$0.07       &  --2.61   & --0.54 \nl
\nl
NI $^f$   & 1134.1653      &  0.01342  &  14.70$\pm$ 0.02 & 10.0 $\pm$ 0.1 & 
3.390123&  --4.03$\pm$0.07   &   --2.68  & --0.61\nl
\nl
NI $^f$  & 1134.4149        & 0.02683  &   &  &         &         &     & \nl  
\nl
NI $^f$   & 1134.9803       & 0.04023   &   &  &         &         &     & \nl  
\nl
OI    & 925.4460 & 0.0003539   & 16.50 $\pm$ 0.03  & 14.3 $\pm$ 1.5 & 3.390161
&   --3.13 $\pm$  0.07      & --1.78    & 0.29 \nl
\nl
OI    & 950.8846        &  0.001571 & 16.35$\pm$0.08  & 10.9$\pm$ 0.9 & 
3.390080 
       &  --3.13 $\pm$  0.07 & --1.93    & 0.14 \nl 
\nl
FeII    & 1611.2004        &  0.001020 $\tablenotemark{c}$& 14.87 $\pm$ 0.03  
& 9.7 $\pm$ 1.3&
3.390105         &   --4.50 $\pm$0.01      & -2.04   & 
0.03\nl
\nl
SiII   & 1808.0126$^f$ &0.002180 & 15.06$\pm$0.02& 9.8 $\pm$ 0.3 & 3.390102& 
--4.44$\pm$0.01        &  --1.91   & 0.16 \nl
\nl
CrII$^f$    & 2056.2539 &   0.1052 $\tablenotemark{d}$&13.09 $\pm$ 0.03 
& 8.8 $\pm$ 0.9 & 3.390078 
       & --6.31 $\pm$0.01       & --2.01   & 0.06\nl
\nl
CrII$^f$    & 2062.2339&    0.07796 $\tablenotemark{d}$&   
&  &  
       &        &    & \nl
\nl
ZnII    & 2026.1360&  0.4886 $\tablenotemark{d}$& 12.01$\pm$0.05 & 
9.3$\pm$1.3&3.390144
& --7.33$\pm$0.04        & --2.07    & -- \nl
\nl
NiII   & 1709.600 &  0.03482 $^{e}$      & 13.39 $\pm$ 0.03 & 8.99 $\pm$ 0.80 
&3.390127 & 
--5.75 $\pm$ 0.01 & --2.27 & --0.20 \nl
\enddata

\tablenotetext{}{ REFERENCES---
 (a) 
Unless otherwise indicated the f values are from Morton (1991);
(b) Prochaska 
\& Wolfe (1999);
(d) Bergeson \& Lawler (1993a);
  (c) Cardelli \& Savage (1995); 
(e) Fedchak \& Lawler (1998);
(f) Bergeson \& Lawler (1993b) }

\tablenotetext{}{NOTES---(f)
The N $\lambda$ 1134 \AA\ was treated as a multiplet in the fitting procedure;
(g)  All abundances were normalised to the meteorite, when available, 
or to photospheric values reported by  Grevesse et al. (1996).}
 
\end{deluxetable}

\clearpage

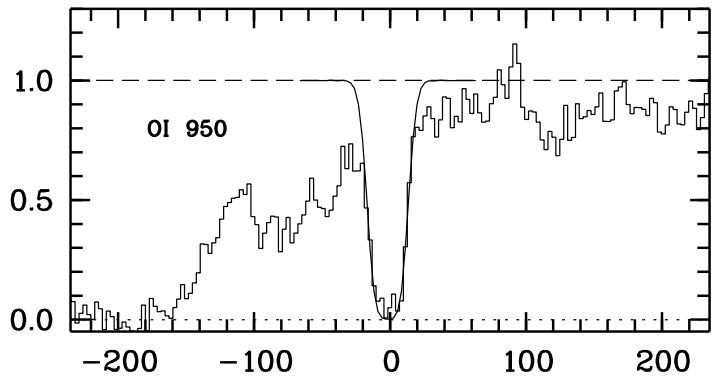
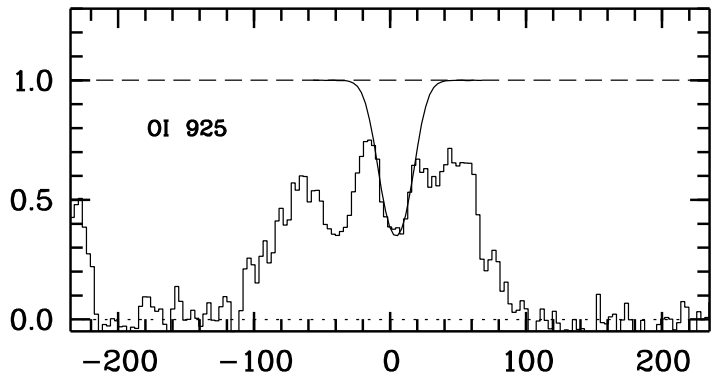
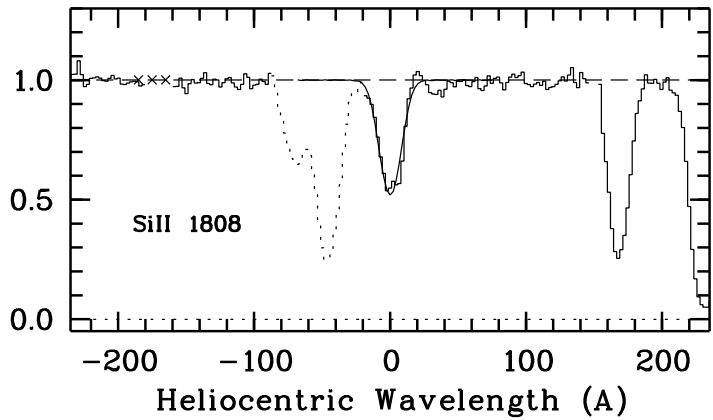
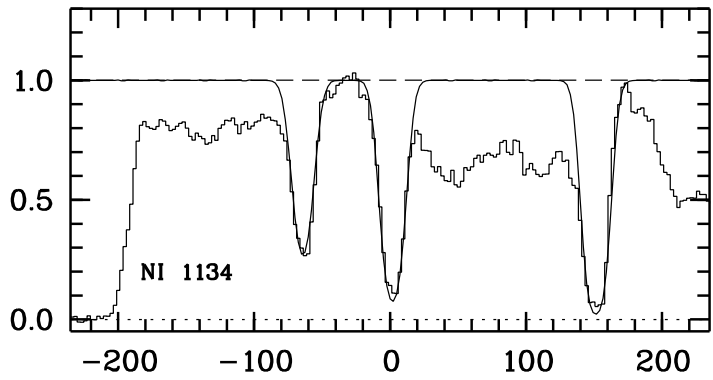
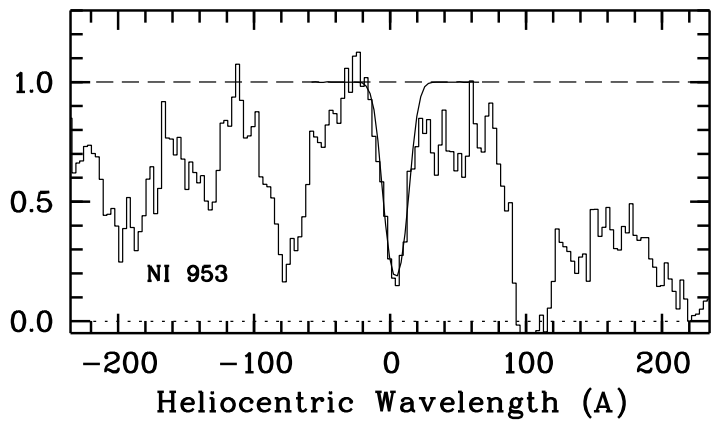
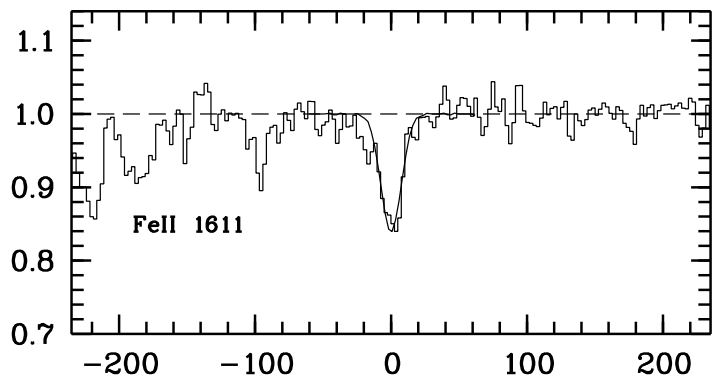
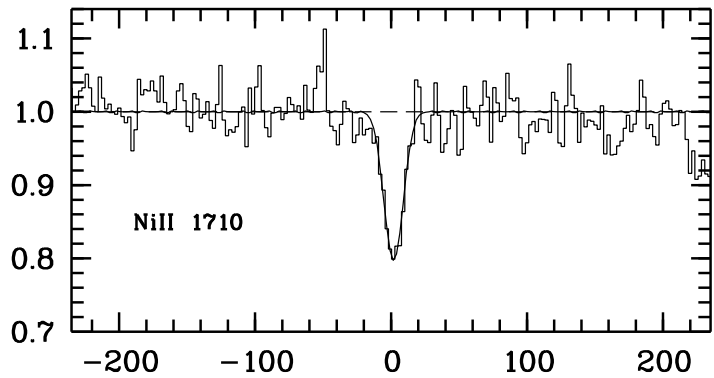
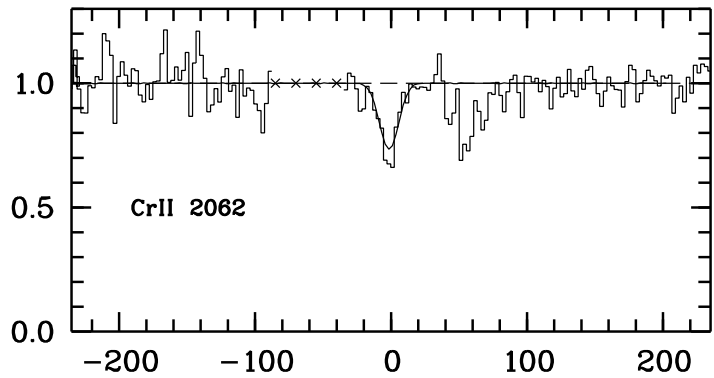
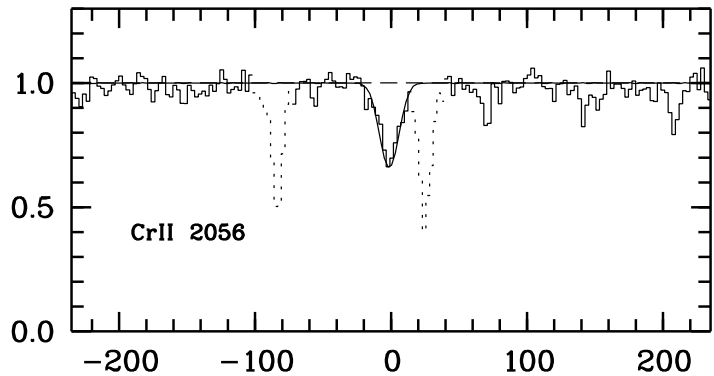
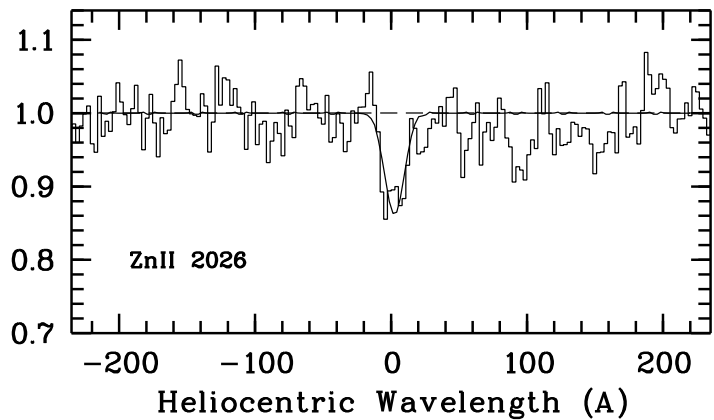
\end{document}